\documentclass{article}
\usepackage{graphicx} 
\usepackage[table]{xcolor}
\usepackage{color, colortbl}

\usepackage{authblk}
\providecommand{\keywords}[1]{\textbf{\textit{Keywords:}} #1}

\usepackage{type1cm}        
\usepackage{makeidx}         
\usepackage{graphicx}        
\usepackage{multicol}        
\usepackage[bottom]{footmisc}

\usepackage{newtxtext}       %
\usepackage[varvw]{newtxmath}       

\makeindex             

\usepackage{color, colortbl}

\usepackage{authblk}
\providecommand{\keywords}[1]{\textbf{\textit{Keywords:}} #1}
\usepackage[pagewise]{lineno}

\usepackage{tabularx,ragged2e,booktabs}
\newcolumntype{L}{%
   >{\RaggedRight\hangafter=1\hangindent=0.667em%
     \textbullet\space}X}

\title{Transforming Triple-Entry Accounting with Machine Learning: A Path to Enhanced Transparency Through Analytics}

\author[1]{Abraham Itzhak Weinberg}
\author[2]{Alessio Faccia}
\affil[1]{Abraham Itzhak Weinberg, AI Experts, Israel, aviw2010@gmail.com}
\affil[2]{ Alessio Faccia, University of Birmingham Dubai, Dubai, UAE, alessio.faccia@gmail.com}

\begin{document}
\maketitle

\begin{abstract}
Triple Entry (TE) is an accounting method that utilizes three accounts or 'entries' to record each transaction, rather than the conventional double-entry bookkeeping system. Existing studies have found that TE accounting, with its additional layer of verification and disclosure of inter-organizational relationships, could help improve transparency in complex financial and supply chain transactions such as blockchain. Machine learning (ML) presents a promising avenue to augment the transparency advantages of TE accounting. By automating some of the data collection and analysis needed for TE bookkeeping, ML techniques have the potential to make this more transparent accounting method scalable for large organizations with complex international supply chains, further enhancing the visibility and trustworthiness of financial reporting. By leveraging ML algorithms, anomalies within distributed ledger data can be swiftly identified, flagging potential instances of fraud or errors. Furthermore, by delving into transaction relationships over time, ML can untangle intricate webs of transactions, shedding light on obscured dealings and adding an investigative dimension. This paper aims to demonstrate the interaction between TE and ML and how they can leverage transparency levels.
\end{abstract}
 \\

\section{Introduction}

\begin{table}
\begin{tabularx}{\textwidth}{@{}LL@{}}
    \toprule
\multicolumn{1}{@{}l}{\textbf{Double-Entry Bookkeeping}} & 
\multicolumn{1}{l@{}}{\textbf{Triple-Entry Accounting}} \\ 
    \midrule
     Records each transaction with two equal and offsetting entries (debits and credits)
    &
     Records each transaction with three entries - the traditional double entries plus an additional digital certificate or record cryptographically recorded on a distributed ledger like blockchain \\ 
    \addlinespace
    Maintains the accounting equation that Assets = Liabilities + Equity
    &
    Provides a single, shared, immutable record on the ledger \\ 
    \addlinespace    
    Entries are recorded privately in a company's own journals and ledgers
    &
    Ledger provides transparency - all entries are visible to permissioned participants on the network \\ 
    \addlinespace   
    Prone to errors or possibility of manipulation as records can be altered
    &
    Reduces risks of fraud, hidden dealings, or manually altered past records \\ 
    \addlinespace
    \multicolumn{1}{l}{} 
    &
    Transactions can be verified in real-time by independent parties on the network \\ 
    \addlinespace
    \multicolumn{1}{l}{} 
    &
    Automated reconciliation simplifies auditing and improves compliance \\ 
    \bottomrule
    \end{tabularx}
    \caption{Comparison of double-entry bookkeeping and triple-entry accounting}
    \label{tab:double_triple_comparison}
\end{table}

Traditional double-entry bookkeeping, a longstanding cornerstone of accounting practices, has historically governed financial record-keeping methodologies \cite{sangster2016genesis}. Its enduring legacy, spanning centuries, has provided a structured framework for tracking financial flows between two entities or accounts. However, as the intricacies of global business transactions and regulatory frameworks have evolved, the inherent limitations of this conventional system have become increasingly pronounced, revealing a lack of the transparency and auditability demanded in contemporary business landscapes. In response to these challenges, the concept of "Triple-Entry (TE) accounting" has emerged as a compelling solution to enrich and modernize accounting practices \cite{tripleentry2022sarin,sgantzos2023triple} as can be seen in Table \ref{tab:double_triple_comparison}.\\ 
In comparing the various models of accounting, it is evident that while traditional and emerging systems aim to enhance the accuracy and transparency of financial records, each model presents distinct differences and limitations. One of the early systems proposed to shift from double entry was the so-called ‘Russian Triple-Entry system’ proposed by Fedor Esersky, which focuses on continuous inventory updates across three books, but its relevance has diminished with the adoption of perpetual inventory systems, making its additional bookkeeping efforts redundant \cite{faccia2020blockchain}. \\
Later, Yuji Ijiri’s Momentum Accounting introduces a third axis, ‘force,’ to capture wealth in addition to income, yet it does not incorporate a third accounting book, limiting its practical application \cite{ijiri1986framework}. More recently, Ian Grigg’s Triple-Entry Accounting integrates cryptography to eliminate reconciliation by using a shared transaction repository \cite{grigg2005triple}. However, this model is largely focused on payment systems and lacks integration with broader accounting principles such as accruals, posing challenges to its widespread adoption; not even the more recent publication from the author has overcome the issue \cite{grigg2024triple}. \\ Another model, the Resource-Event-Agent (REA) proposed by McCarthy, aims to enable a computer software model for an enterprise information system by focusing on inter-enterprise shared ledger systems. However, it replaces rather than integrates with existing AIS/ERP systems, limiting its compatibility with current infrastructures \cite{mccarthy1982rea}. The X-Accounting model, developed by Faccia, advances these concepts by integrating blockchain technology and AI to create a ‘Triple-Axis’ system that enhances accountability and real-time auditing \cite{faccia2020blockchain}. \\
Despite its potential, this model requires significant infrastructure upgrades and government enforcement, highlighting the complexity of fully implementing such advanced systems in existing accounting frameworks. Triple-Axis X-Accounting is an advanced accounting model that enhances traditional double-entry bookkeeping by adding a third axis: a blockchain-generated hash. This model records each transaction in three ways: Debit and Credit Entries, which is the traditional method of recording transactions in two accounts. Blockchain Hash: A cryptographic hash unique to each transaction, ensuring it is securely recorded and immutable. AI-Driven Adjustments: Automation through artificial intelligence for real-time auditing, compliance checks, and adjusting entries. This third axis (the blockchain hash) provides a tamper-proof, time-stamped record, making each transaction traceable and transparent across all participating entities.\\ The use of AI further streamlines the accounting process by automating routine tasks and enabling continuous auditing. X-Accounting works by integrating these components into a shared ledger accessible by all relevant parties, ensuring that all records are consistent, secure, and up-to-date. This model not only improves the security and accuracy of financial records but also facilitates real-time compliance and scalability, making it an efficient and modern solution for complex accounting needs \cite{faccia2020blockchain,faccia2019accounting,faccia2021financial}.\\
By integrating a third entry alongside the traditional debits and credits, TE accounting extends the narrative of financial transactions by encapsulating non-financial contextual data, elucidating the comprehensive who, what, where, when, and why behind each financial exchange. This innovative approach not only enhances transparency by rendering financial decisions and their underlying rationale auditable but also fortifies defenses against fraudulent activities seeking to obscure transactional specifics \cite{grigg2005triple}. Furthermore, the inclusion of supplementary metadata fosters opportunities for advanced analytics to extract deeper insights into organizational performance, risk management, and operational efficiencies. In an era characterized by a "trust but verify" ethos, the enhanced transparency and evidence-based nature of TE accounting serve to bolster credibility with critical external stakeholders, including investors, regulators, and auditors \cite{groblacher2019triple}. \\
Nevertheless, the mere accumulation and storage of enriched transactional data are insufficient without systematic analysis through the lens of advanced analytics. Here, the integration of ML emerges as a pivotal enabler, automating the detection of patterns, relationships, and anomalies across both financial and contextual dimensions. This infusion of data science techniques into oversight processes not only elevates transparency and compliance efforts but also propels strategic decision-making to unprecedented levels, equipping organizations to navigate the multifaceted challenges of today's intricate business environment with newfound clarity and agility \cite{chowdhury2021financial}.\\
While the idea of TE accounting aims to increase transparency, its full potential can only be achieved by leveraging advanced analytics on the richer transactional data \cite{petratos2024triple,ibanez2023rea}. Capturing additional metadata through a third accounting entry addresses only part of the challenge - the real value comes from what organizations can uncover by systematically analyzing these combined financial and contextual records.
Traditional audit and reporting practices may not be sufficient for making full sense of the large volumes and complexity inherent in TE accounting data. This is where integrating ML techniques can help automate the process of discovery \cite{dixit2024incorporating}. \\
For example, classification algorithms could help identify common patterns and anomalies across attributes like customer segments, product categories, geographical regions or time periods. Clustering approaches may group entities or transactions based on similarities in their multiple dimensions. These analytical insights can then shine a light on areas warranting deeper scrutiny, as inconsistencies or outliers come to the surface. \\
Another benefit of the analytics layer is connecting seemingly disparate pieces of information \cite{qadir2022blockchain}. For instance, correlating specific contextual attributes with downstream financial metrics could reveal important drivers of profitability, revenue or costs that may not be apparent simply through manual examination. Such enriched associative links between various transactional attributes are what truly drive strategic decision-making value. Advanced algorithms are well-equipped to unveil these subtle yet meaningful multivariate relationships buried in vast volumes of interlinked data.
Transparency is crucial for trust, accountability, security and true decentralization in Decentralized Finance (DeFi) systems such as blockchains since it helps promote adoption and integrity of these new open financial networks. TE accounting fills in the gaps of DeFi and blockchains \cite{schuldt2023decentralized}. The transparency, immutability and verifiability of TE accounting perfectly complements the decentralized nature of DeFi applications and blockchains. It provides essential accounting functionality for this new digital financial system.

\subsection{Triple Entry Accounting and Smart Contracts}
Within TE accounting, the connection with smart contracts is crucial for various aspects. \cite{faccia2019accounting}. A smart contract is defined as transaction protocol designed to autonomously execute, govern, or record events and actions based on the conditions of a contract or agreement \cite{zou2019smart}. TE accounting utilizes a distributed ledger, typically a blockchain, to capture the third entry, guaranteeing an unchangeable transaction history \cite{inder2023triple}. The comparison between TE and smart contracts can be seen in Table ~\ref{tab:smertcn_triple_comparison}. \\
In parallel, smart contracts empower the automation of transaction conditions and executions, executing contractual clauses that self-verify and self-execute. This automation cultivates trustless verification, removing the necessity for intermediaries and heightening transparency. Moreover, by embedding terms in smart contracts, the automation of reconciling TE bookkeeping entries reduces manual errors and labor. The secure storage of records within smart contracts on a distributed ledger ensures the trustworthiness of financial reporting. Auditors gain from streamlined auditing processes by inspecting immutable ledger records generated through smart contract executions. Furthermore, real-time updates facilitated by smart contracts amplify visibility and enable instant record reconciliation. Conditional payments delineated in smart contracts ensure the automatic fulfillment of payment obligations upon meeting predefined conditions, thereby optimizing processes effectively.

\begin{table}[ht]
\centering
\begin{tabular}{|l|p{6cm}|p{6cm}|}
\hline
\rowcolor{blue!30}
& \textbf{Triple-Entry Accounting} & \textbf{Smart Contracts} \\ \hline
Distributed Ledger & Can use any distributed ledger (blockchain, 
Directed Acyclic Graph) \cite{schuldt2023decentralized} & Requires a blockchain or smart contract platform \\ \hline
Verification & Transactions verified by participants & Transactions verified by network consensus \\ \hline
Immutability & Ledger entries are immutable & Code and execution are immutable \\ \hline
Transparency & Transactions are transparent & Code, execution, and state are transparent \\ \hline
Automation & Manual accounting processes & Processes are automated through code \\ \hline
Conditional Actions & No conditional logic & Include conditional logic through code \\ \hline
Governance & Requires trusted central authorities & Fully decentralized governance \\ \hline
Fraud Prevention & Reduces alteration risk & Reduces alteration and misconduct risk \\ \hline
Auditing & Simplifies auditing process & Enables programmatic auditing of execution \\ \hline
Limitations & Privacy, scaling, and interoperability & Bugs, centralization of platforms \\ \hline
\end{tabular}
\caption{Comparison of Triple-Entry Accounting and Smart Contracts}
\label{tab:smertcn_triple_comparison}
\end{table}

\subsection{The Role of Triple-Entry Accounting and Alternatives}
TE accounting is regarded as a crucial advancement for several reasons. It enhances transparency and trust by documenting financial transactions from three distinct viewpoints: the involved parties and an impartial third ledger, fostering greater confidence in the accuracy of financial records. Moreover, it enables a comprehensive traceability of value flow, surpassing the limitations of traditional double-entry systems by tracking how value circulates through interconnected transactions, yielding deeper insights. \\ Additionally, its compatibility with distributed ledger technologies like blockchain facilitates independent verification of transactions without a central authority, aligning well with principles of transparency. Furthermore, TE accounting extends to sustainability considerations, encompassing natural, social, and relationship capital alongside monetary flows, potentially aiding in environmental, social, and governance reporting for overall sustainability impact assessment. Lastly, it serves as a robust tool for mitigating fraud and errors, with the third ledger acting as a vigilant checkpoint to identify inconsistencies or inaccuracies, thereby reducing mistakes and fortifying fraud prevention mechanisms.

\subsection{Pros and Cons of Triple-Entry Accounting}
TE accounting offers several advantages and disadvantages worth considering as can be seen in Table \ref{tab:proscons}. On the positive side, it enhances transparency and auditability by introducing an additional verification layer, improving accuracy and early detection of errors or fraud through easily identifiable ledger inconsistencies. 
\begin{table}
\centering
\begin{tabular}{|p{7cm}|p{7cm}|}
\hline
\rowcolor{blue!30}
\textbf{Pros} & \textbf{Cons} \\
\hline
Increased transparency and auditability of financial records. & Higher compliance costs compared to traditional double-entry bookkeeping. \\
Improved accuracy and ability to detect errors or fraud early. & Increased complexity of financial record-keeping systems and processes. \\
Better suited for distributed ledger technologies due to independent verification. & Potential issues synchronizing updates across all three ledgers in real-time. \\
Standardizes capturing of multi-step transactions and non-financial impacts. & Privacy and confidentiality may be reduced with full transparency of records. \\
Facilitates tracking value flows through entire systems of interconnected parties. & Immaturity of standards and lack of experience operating TE systems in practice. \\
Promotes trust in accounting through consensus-based self-verification. & Reporting frameworks still evolving to translate additional data captured into meaningful reporting. \\
Might be overkill for simpler organizations or transactions that don't require full transparency. & \\
\hline
\end{tabular}
\caption{Pros and Cons of Triple-Entry Accounting}
\label{tab:proscons}
\end{table}

It is well-suited for distributed ledger technologies, standardizes the recording of complex transactions, and enables the tracking of value flows across interconnected systems. Moreover, it fosters trust in accounting practices by promoting consensus-based self-verification. Conversely, adopting TE accounting can incur higher compliance costs, complicate financial record-keeping processes, and pose challenges in synchronizing updates across all three ledgers in real-time. The full transparency of records may reduce privacy and confidentiality, while the immaturity of standards and lack of operational experience with TE systems present additional hurdles. Furthermore, reporting frameworks are still evolving to effectively translate the captured data into meaningful reports, and for simpler organizations or transactions not requiring full transparency, TE accounting may be excessive.\\  In essence, while TE accounting enhances transparency, integrity, and interoperability \footnote{Blockchain interoperability allows for different blockchain systems to communicate and exchange data, messages, and digital assets with one another.}, it also introduces complexities, costs, and potential synchronization and privacy issues compared to traditional methods, with standards and practical experience still in development.

\subsection{Why is Triple-Entry Accounting so Important for Blockchain?}
TE accounting holds immense significance for blockchain due to its intrinsic alignment with the decentralized nature of blockchain technology \cite{faccia2019accounting}. By incorporating a third independent ledger to verify transactions, TE accounting enhances transparency and trust within blockchain networks, crucial for ensuring the accuracy and integrity of financial records. This additional layer of verification not only mitigates fraud and errors but also complements the decentralized and distributed nature of blockchain by providing a reliable mechanism for validating transactions across the network without the need for a central authority. Furthermore, the traceability of value flow inherent in TE accounting complements blockchain's ability to securely track and record the movement of assets, making it an essential tool for enhancing the efficiency, security, and reliability of blockchain transactions.

\subsection{The Importance of Transparency in Blockchain Technology}
Transparency plays a critical role in blockchain technologies for several key reasons \cite{cai2021triple}. In decentralized blockchain systems where a central authority is absent, transparency is essential to enable all participants to independently verify the validity and sequencing of transactions, fostering a trustless environment where reliance on third parties is unnecessary. The immutability characteristic of blockchains, aiming to maintain an unchangeable record, relies on transparency into rule-based operations and comprehensive transaction histories for users to continuously validate the integrity and accuracy of the ledger.\\ Moreover, the auditability aspect is crucial for regulators and auditors who seek assurance that blockchains function as transparent public ledgers, facilitating asset tracking, error and fraud detection, and compliance monitoring. Additionally, transparency in decentralized blockchains ensures censorship resistance by making data equally visible to all users, unlike centralized systems vulnerable to data manipulation or concealment. Furthermore, in blockchain protocols like Bitcoin that operate without trusted third parties, transparency into the consensus protocol and transaction confirmation process is vital for users to ascertain the finality of transactions. This transparency also fosters accountability by making all data, algorithms, and transaction processes open for public scrutiny, incentivizing protocols to operate with integrity and minimizing the risk of unintended flaws or manipulations that could erode users' trust in the system.

\subsection{Alternatives for Transparency in Accounting and Blockchain Systems}
In exploring alternatives to transparency in accounting, several avenues diverge from the conventional norm of full disclosure \cite{alkan2021real}. Compliance reporting focuses on meeting regulatory requirements with minimal information, reducing transparency by providing only essential data. Private or restricted ledgers keep accounting records confidential, diminishing visibility and trust due to the absence of independent verification on public distributed ledgers. Estimates and approximations involve using rounded figures and aggregated data rather than precise numbers, offering reduced transparency while potentially sufficing for certain applications. \\ 
Selective disclosure entails releasing high-level metrics instead of granular transaction data, limiting transparency scope to streamline information dissemination. Third-party attestation involves independent auditors verifying accuracy instead of complete transparency on distributed ledgers, introducing a trusted intermediary \cite{dai2017toward}. Proprietary formats store records in closed formats, hindering transparency and impeding interaction for independent participants. Manual reconciliation, involving periodic rather than real-time balance checks, lowers the frequency and reliability of transparency. Anonymized metadata shares transaction details without identifiers, reducing linkability and contextual clarity compared to full transparency of counterparties and items.\\
When considering alternatives to TE accounting within blockchain systems, various approaches emerge that offer different paths towards transaction verification and accuracy \cite{bonson2019blockchain,fullana2021accounting,rizal2019unraveling}. Cryptographic proofs, such as zero-knowledge proofs \footnote{A zero-knowledge proof or protocol is a technique through which one party can demonstrate to another party the truth of a given statement without revealing any details beyond the validity of that statement.}, cryptographically validate transaction accuracy without exposing full transaction details, providing verification without complete transparency. Auditing committees could replace the TE system by periodically selecting committees to review transactions, adding trusted third-party auditors into the verification process. Off-chain data availability involves storing transaction payloads off-chain while maintaining light client-verifiable commitments on-chain, enabling verification through sampling rather than for each transaction.\\
Selective disclosure focuses on releasing high-level summaries instead of granular transaction details, emphasizing transparency through analytics over raw data. Privacy technologies like homomorphic encryption \footnote{Homomorphic encryption is a cryptographic method enabling computations to be conducted on encrypted data without the need for decryption.} computations on private data without revealing content, offering verification without accessing raw data \cite{shrestha2019integration}. \\
Blockchain sampling validates random blocks periodically instead of real-time validation for every transaction, reducing overhead at the expense of real-time assurance \cite{elommal2021blockchain}. Despite these alternatives, most compromise transparency to some extent or introduce additional parties and assumptions compared to the self-auditing nature of TE accounting. Thus, TE accounting remains the most direct approach to achieve transparency, integrity, and decentralization simultaneously within blockchain systems.

\section{Foundations of Triple-Entry Accounting}
Double-entry bookkeeping, a foundational practice within accounting, has long been relied upon for maintaining financial records with precision \cite{cai2021triple}. This system's integrity is upheld by ensuring that debits align with credits and transactions are recorded twice, instilling consistency in financial documentation \cite{rahmawati2023demistifying}. Although double-entry bookkeeping excels in balancing accounts, it falls short in providing a comprehensive understanding of transaction contexts, such as the parties involved, the nature of exchanges, and the timing and rationale behind transactions.\\
In response to this limitation, the principles of TE accounting have been introduced to augment traditional practices \cite{cai2021triple}. This innovative approach integrates a third "entry" dedicated to capturing non-financial metadata alongside debit and credit entries. By intertwining financial data with crucial contextual details, TE accounting offers a more holistic view of organizational activities, propelling insights beyond those achievable through traditional double-entry systems. This rich contextual foundation serves as a robust platform for leveraging advanced analytics methodologies, enabling organizations to derive deeper insights and make well-informed strategic decisions based on a wealth of interconnected data points.

\subsection{Limitations of Traditional Double-Entry Bookkeeping}
Traditional double-entry bookkeeping, while a stalwart in accounting practices, grapples with several key limitations that TE accounting seeks to rectify \cite{chowdhury2021financial,hambiralovic2018blockchain}. The current system lacks transparency as it focuses solely on monetary exchanges between accounts, omitting vital contextual details crucial for verifying transactions and detecting anomalies. Furthermore, the inability to connect financial data with non-financial information creates a disjointed perspective on transactions, impeding comprehensive analysis and audit trails. \\
This deficiency not only obscures potential fraud and errors but also complicates compliance adherence, as the absence of transaction context can hinder regulatory assessments. Moreover, the limited viewpoint provided by the traditional double-entry system restricts strategic analysis by overlooking interconnected attributes like location, product, and customer data, which are essential for obtaining deeper operational insights. Finally, the system's inability to trace asset provenance across successive transactions undermines the clarity and integrity of historical financial records, weakening independent verification and heightening the risk of unnoticed anomalies.

\subsection{Overview of Triple-Entry Accounting: Principles and Framework}
The principles and framework of TE accounting represent a significant advancement from traditional double-entry systems \cite{groblacher2019triple,rahmawati2023demistifying}. By expanding beyond the financial debits and credits paradigm, TE accounting introduces a third "entry" dedicated to capturing contextual metadata for each transaction, offering a comprehensive view of the "why" behind financial flows \cite{tripleentry2022sarin}. \\
In this framework, financial elements are recorded conventionally as debits and credits, while non-financial attributes such as parties involved, transaction locations, item descriptions, and tags are meticulously defined as metadata fields. Each transaction is uniquely identifiable through shared reference keys, ensuring a robust linkage between financial and contextual entries that are both immutable and transparent. This model not only bolsters auditability and transparency but also facilitates deeper insights and analytics by providing a unified and verifiable view of both financial and contextual activities. By enabling the seamless integration of financial and non-financial data in a structured and immutable manner, TE accounting not only fortifies compliance and audit processes but also serves as a resilient foundation for advanced analytics approaches within accounting practices.
In TE Accounting, a concept that enhances traditional accounting by introducing a third entry to the transaction, there are specific operational mechanisms that define its implementation. This innovative framework involves not only the debit and credit entries found in double-entry accounting but also a third entry that is cryptographically sealed to provide an additional layer of transparency and security.\\
To delve deeper into the integration of non-financial metadata within this system, it is essential to elaborate on the process through which this data is captured, stored, and linked to financial transactions. Non-financial metadata can encompass a wide range of information, such as timestamps, transaction details, or even external data like contracts or agreements.\\
Non-financial metadata plays a vital role in enhancing the understanding of financial transactions within TE Accounting \cite{spilnyk2020accounting}. This data, captured during transactions, offers essential context to comprehend the significance and implications of financial entries. Various methods, such as manual input, automated data feeds, or integration with external systems, are employed to collect this metadata, ensuring a comprehensive view of each transaction's context and purpose.\\
In TE Accounting, the secure storage of non-financial metadata alongside financial data is paramount for maintaining a comprehensive transaction record \cite{sgantzos2023triple}. This practice ensures that each transaction is richly documented with contextual information, offering a deeper understanding of the financial entries. By leveraging secure databases or blockchain technology, the integrity and immutability of this combined dataset are safeguarded, providing a reliable and transparent record of the transactional history.\\
The linkage between financial transactions and their associated non-financial metadata is pivotal for constructing a comprehensive understanding of the transactional context and history \cite{sunde2023implementing}. By establishing a connection between each financial entry and its pertinent non-financial metadata, a holistic perspective is formed, shedding light on the nuances and implications of the transactions. This linkage is facilitated through the use of unique identifiers or cryptographic hashes, serving as the glue that binds together the financial and non-financial components, ensuring a cohesive and transparent representation of the transactional data.\\
By incorporating these detailed explanations of the operational mechanisms involved in capturing, storing, and linking non-financial metadata within the TE Accounting framework, a more comprehensive understanding of its practical application and benefits can be attained.

\section{The Role of Advanced Analytics in Triple-Entry Accounting}
\label{sec:2}
While TE accounting enriches transactional insights by interlinking financial and metadata records, the mere accumulation of this comprehensive data is insufficient without systematic analysis and value extraction mechanisms. Organizations necessitate automated and intelligent processes to sift through vast amounts of information effectively. This is where the integration of advanced analytics driven by ML emerges as crucial. Traditional manual auditing and reporting may overlook intricate patterns, correlations, and anomalies embedded within multifaceted datasets, whereas ML algorithms excel in uncovering hidden insights and knowledge from complex data structures. \\ Techniques like prediction, classification, clustering, and association rule mining can be harnessed to unearth strategic and tactical insights, enabling organizations to forecast future outcomes, identify anomalies, segment entities, discover relationships, and optimize processes. By automating compliance monitoring, enhancing fraud detection, and facilitating informed decision-making, this analytical fusion empowers auditors, managers, and regulators with scalable tools for continuous oversight, strategic planning, and assurance of organizational integrity. By leveraging the cognitive capabilities of ML within transparent TE accounting frameworks, a new era of analytical prowess emerges, promising to redefine accountability, governance, and value generation across industries navigating the complexities of today's data-driven business landscape.

\subsection{Motivation for Applying Machine Learning and Data Mining}
The integration of ML and data mining into TE accounting data unveils a spectrum of compelling reasons driving this technological fusion. ML's prowess lies in unearthing obscured patterns and relationships within vast and diverse transaction datasets, offering insights beyond human perception. Predictive models, a hallmark of ML, furnish a methodical approach to foresee outcomes, predict risks, identify anomalies, and estimate future financial trajectories based on historical metadata trends. By automating monitoring processes and anomaly detection, ML algorithms provide continuous oversight, swiftly flagging irregularities and conserving manual efforts. This proactive monitoring extends to auditing and governance realms, ensuring real-time alignment with policies rather than sporadic assessments, fostering a culture of perpetual compliance adherence. \\
Moreover, ML generated patterns and predictive guidance equip managers with evidence-based support for strategic decision-making across various organizational domains. Addressing the challenges posed by extensive data volumes inherent in TE accounting, ML excels in extracting intelligence from interconnected datasets at a scale surpassing traditional tools. \\ Additionally, ML classifiers and rules play a pivotal role in streamlining compliance efforts, systematically checking for violations and mitigating oversight costs associated with manual reviews. By spotlighting outliers and enhancing forensic capabilities, ML becomes instrumental in combating fraud and errors, bringing to light both intentional and unintentional financial reporting discrepancies. Furthermore, the iterative learning process facilitated by ML fosters organizational knowledge enhancement over time, with insights garnered from feedback loops contributing to the model's accuracy as it delves into recurring and novel transactions alike.\\
In the context of TE Accounting, the selection of ML algorithms is paramount, with certain types, such as clustering and classification algorithms, holding particular relevance due to their distinct characteristics and capabilities.
Clustering algorithms, like K-means or hierarchical clustering, are well-suited for Triple Entry data due to their ability to identify patterns and group transactions based on similarities \cite{centorrino2022double}. This feature is invaluable for anomaly detection within financial data, allowing for the detection of irregularities or potential fraudulent activities within the transaction records.\\
On the other hand, classification algorithms, such as decision trees or Support Vector Machines (SVM), play a vital role in TE Accounting by enabling the categorization of transactions into distinct classes based on historical patterns and data features \cite{dixit2024incorporating}. These algorithms can help in assigning labels to transactions, such as legitimate or suspicious, thus aiding in the identification of potentially fraudulent transactions or unusual patterns that deviate from the norm.\\
Moreover, the interpretability \footnote{Interpretability refers to the degree to which cause-and-effect relationships can be discerned within a system. Put differently, it reflects the extent to which one can anticipate outcomes following modifications in input or algorithmic parameters \cite{murdoch2019definitions}.} and explainability \footnote{Explainability means clarifying how a model progresses from input to output, enhancing transparency and addressing the black box issue. This concept, formalized as Explainable AI (XAI), applies universally to artificial intelligence, focusing on making AI systems interpretable and accessible \cite{burkart2021survey}.} of these algorithms are crucial in the financial domain, as they provide insights into the reasoning behind the classification or clustering decisions, which is essential for auditing and compliance purposes \cite{zhang2022explainable}. By leveraging clustering and classification algorithms in Triple Entry Accounting, organizations can enhance their ability to analyze and interpret financial data effectively, leading to improved risk management, fraud detection, and decision-making processes within the realm of financial transactions.

\subsection{Benefits of Transparency, Insights and Fraud Detection}
Empowering TE accounting with ML yields a myriad of advantages spanning transparency, insights, and fraud detection. By integrating ML into the accounting framework, visibility into full transaction context is enhanced through the interconnectedness of financial and metadata records, with ML algorithms further enhancing transparency by automatically identifying anomalies and outliers \cite{awosika2024transparency}. This fusion also fortifies compliance and governance efforts through real-time monitoring, alerts, and the application of classification and rules to pinpoint regulatory issues at a detailed level. \\
Strategic decision-making gains a competitive edge as data-driven predictive patterns and associations offer objective guidance, shedding light on optimization opportunities across various dimensions. The process of continuous organizational learning is facilitated through feedback loops, refining models and knowledge over iteratively analyzed data, thereby influencing well-informed planning and decision-making over the long term. \\
Moreover, automated fraud and error detection mechanisms enable the identification of unusual profiles and transactions beyond set thresholds, reducing oversight costs while bolstering forensic investigation support. Proactive risk prediction is enabled through predictive models that forecast potential risks, liabilities, and problematic areas, allowing for preemptive controls to be implemented, thus strengthening risk mitigation strategies compared to reactive post-event approaches. Ultimately, this amalgamation of TE accounting with ML not only enhances credibility for stakeholders by providing intelligent oversight but also minimizes disruptions stemming from compliance issues or reporting inaccuracies, solidifying the strategic benefits across governance, compliance, and decision-making functions within organizations.

\section{A Machine Learning Approach}
The methodology for representing TE data as features involves converting financial records into target or output variables, while metadata attributes are reshaped into predictive input features. Categorical variables are managed through dimensional modeling techniques, and data irregularities are addressed using feature encoding methods \cite{cho2020learning,ucoglu2020current}. Various algorithms and models are then employed for different purposes: classification models such as logistic regression \cite{bekoe2018attitudes}, decision trees \cite{nawaiseh2021financial}, and neural networks \cite{sanchez2020predicting} are utilized to predict transaction types or outcomes, while clustering algorithms like K-means \cite{sahoo2021accounting}, hierarchical clustering \cite{thiprungsri2019cluster}, and Density-Based Spatial Clustering of Applications with Noise (DBSCAN) \cite{bhattacharya2020semi} are used to identify natural segments within similar records. \\
Association rule mining techniques \cite{sawangarreerak2021detecting} like Apriori \cite{ioannou2021framework} and Equivalence CLAss Transformation (ECLAT) \cite{tsapani2020knowledge} are applied to uncover correlations, and anomaly detection algorithms including Isolation Forests \cite{bakumenko2022detecting}, Local Outlier Factor (LOF) \cite{kuna2014outlier}, and One Class Support Vector Machine (OCSVM) \cite{wilfred2021big} are employed to detect outliers. \\ Predictive modeling methods like linear or logistic regression \cite{nawaiseh2021financial}, Recurrent Neural Networks (RNNs) \cite{jan2021using}, and time-series analysis are utilized for forecasting purposes. Model evaluation is carried out using realistic train/test datasets, with performance metrics tailored to each specific problem, such as accuracy, precision, recall for classification tasks \cite{nawaiseh2021financial}, silhouette score for clustering \cite{byrnes2019automated}, confidence and support for associations \cite{suppiah2023impact}, and F-score \cite{miharsi2024analysis}, Receiver Operating Characteristic (ROC) Area Under the Curve (AUC) \cite{carrington2022deep} for anomalies, with strategies in place to combat overfitting and data imbalances. This systematic approach aims to transform TE data into actionable intelligence using a diverse array of ML algorithms, tailored to the unique analytical objectives at hand.

\subsection{Methodology for Representing Triple-Entry Data as Features}
In the methodology for representing TE data as features, financial records are designated as target or label variables, with debit and credit amounts potentially being predicted or classified as the output or label, while fields such as account and transaction ID serve as unique identifiers \cite{dixit2024incorporating,ibanez2021efficiency}. Metadata attributes \footnote{attributes that describes data.} are converted into predictive features, with categorical variables like location encoded using techniques such as one-hot encoding \footnote{A technique that converts categorical data into numerical by creating binary columns for each category, marking 1 for presence and 0 for absence.}, dates transformed into integer time deltas or periods encoded using one-hot encoding, and text fields processed using methods like TF-IDF \footnote{an acronym for Term Frequency-Inverse Document Frequency, quantifies the relevance of a word in a text by comparing its frequency across a corpus.} or word embeddings \footnote{represents a word as a real-valued vector, encoding its meaning such that words closer in the vector space are likely to share similar meanings.}.\\
Dimensional modeling manages categorical variables by creating separate tables for textual fields like product or customer, extracting categorical variables from these dimension tables, and employing one-hot encoding \cite{sarwar2023blockchains}. Feature engineering \footnote{A supervised ML processing, where raw data is converted into a more efficient set of inputs, with each input consisting of multiple attributes, referred to as features.} plays a crucial role in structuring inputs appropriately, involving the grouping of related fields into composite features when meaningful, normalization to standardize numeric ranges \footnote{A scaling technique used to standardize the values of numeric columns in the dataset to a uniform scale.} like debit amounts, and handling missing values or sparse attributes. Feature selection techniques are utilized for dimensionality reduction \footnote{The process of reducing data from a high-dimensional space to a low-dimensional space while preserving key properties of the original data.}, identifying unimportant or correlated inputs through methods like ANalysis Of VAriance (ANOVA)\cite{surana2022awareness} \footnote{A test employed to ascertain variances between research findings across three or more distinct and unrelated samples or groups.} and employing recursive feature elimination to enhance algorithm performance. This structured approach enables ML algorithms to uncover valuable patterns and relationships within TE data, facilitating tasks such as classification, clustering, and prediction for various analytical objectives.

\section{The Relation between Triple-Entry Accounting and Multiparty Computation (MPC)}
\begin{table}[ht]
\centering
\begin{tabular}{|p{3cm}|p{4cm}|p{4cm}|}
\hline
\rowcolor{blue!30}
& \textbf{Triple-Entry Accounting} & \textbf{Multiparty Computation} \\
\hline\hline
Approach & Links financial transactions to contextual metadata and stores permanently & Computes on private inputs without directly revealing them through secure sharing \\
\hline
Scope & Focuses on accounting domains and transparency & General technique for private joint analysis by multiple parties \\
\hline
Data Management & Collects and archives all records & Computes without long-term storage of private inputs/outputs \\
\hline
Privacy & Provides transparency of record contents & Preserves privacy of inputs during distributed computation \\
\hline
Implementation & Leverages database technologies & Relies on cryptographic secure computation protocols \\
\hline
Use Cases & Auditing, compliance, analytics on public records & Collaborative analysis of sensitive private data \\
\hline
Strengths & Allows archiving and transparency & Preserves privacy during joint analysis \\
\hline
Weaknesses & No long-term input privacy & No permanent archive of results \\
\hline
\end{tabular}
\caption{Comparison of Triple-Entry Accounting and Multiparty Computation for Privacy-Preserving Analysis of Transaction Data}
\label{tab:tec_mpc_comparison}
\end{table}
Multiparty Computation (MPC) also known as protocol of oblivious transfer objective is to develop techniques that enable parties to collaboratively compute a function using their inputs while maintaining the privacy of those inputs \cite{evans2018pragmatic}. MPC fits well with blockchains by enabling privacy-preserving and decentralized computations in a transparent and incentive-compatible manner - all of which are important requirements for many blockchain use cases \cite{zhou2021using}. TE accounting and MPC can be complementary technologies. TE alone cannot replace the cryptographic security guarantees provided by MPC-based solutions for multiparty computations on blockchains as can be seen in Table \ref{tab:tec_mpc_comparison}. MPC is still needed for that core functionality \cite{ibanez2021efficiency,mccallig2019establishing,cao2024distributed}.\\
The differences between the two lie in their core functionalities: MPC stands out as a groundbreaking method facilitating secure computations among multiple parties without exposing their private inputs, ensuring a high level of confidentiality. This technique, underpinned by robust cryptographic principles, offers security assurances even in scenarios where some parties may collude or act maliciously. \\ In contrast, TE Accounting, a system tailored for the immutable and auditable recording of financial transactions, lacks the same comprehensive security guarantees intrinsic to MPC. While TE Accounting excels in maintaining transaction records on a decentralized ledger, it falls short in enabling secure distributed computations akin to the capabilities provided by MPC.\\
They can be combined to create a robust framework: MPC protocols offer a secure avenue for computing transactions and business logic, with the outcomes subsequently documented on-chain \footnote{A transaction that is exclusively registered and validated within the primary blockchain.} in a triple entry format for enhanced transparency. Integrating zero-knowledge proofs \cite{sun2021survey}, generated via MPC, with TE accounting could furnish mathematical verification of transaction validity. \\ Moreover, on-chain MPC could uphold regulatory compliance logic, channeling the results into TE accounting systems for comprehensive governance and oversight.\\
MPC and TE accounting possess complementary strengths: MPC excels in secure computations, while TE accounting ensures transparency. By integrating the two, MPC can manage confidential on-chain computations, while TE accounting securely records the outputs in an immutable manner, leveraging the strengths of both systems for a comprehensive and efficient solution.

\subsection{The Potential of Integrating Triple Entry Accounting with Multi-Party Computation (MPC)}
MPC can significantly enhance triple entry accounting on blockchain networks through various applications such as privacy-preserving audits and compliance checks, securing records linkage, regulatory calculations, fraud detection, dispute resolution, running encrypted queries and risk modeling.\\
Privacy-preserving audits and compliance checks can be conducted using MPC, allowing sensitive financial data such as transactions and Know Your Client (KYC) records \footnote{KYC, short for Know Your Customer or Know Your Client, is the mandatory process where banks verify clients' identities to ensure authenticity.} to undergo auditing or compliance logic checks without disclosing the underlying information. The pass/fail results can then be securely recorded in the TE system.  \\ MPC also enables secure record linkage, facilitating the linking of related records like transactions across different parties or systems in a privacy-preserving manner. This linkage supports functionalities such as netting and reconciliation within the TE logs. Furthermore, MPC can power fraud detection algorithms, enabling the private execution of complex ML/AI models on transaction data to identify issues, with the findings reported anonymously in the triple entry format. \\
Regulatory calculations, including determining prudential standards and capital adequacy ratios, can be privately computed using MPC on financial data before public reporting. In the event of disputes, MPC protocols can be employed to verify facts or compute ledger states privately, aiding in dispute resolution while minimizing data disclosure. \\ 
Users can utilize encrypted queries through MPC to run queries on the triple entry ledger, extracting private analytics without revealing the queries or results. Lastly, for risk modeling purposes, MPC can be leveraged to analyze counterparty risk measures, conduct stress tests, and perform portfolio analyses privately, enhancing risk assessment across the network.\\
As previously stated, MPC plays a crucial role in transforming privacy-preserving audits in the domain of TE accounting by providing secure means to verify transactions and maintain compliance without jeopardizing sensitive financial information. For instance, a consortium of financial institutions adopts MPC protocols to conduct audit trail verifications securely, enabling the comprehensive validation of transaction histories across multiple parties without revealing individual transaction details. Similarly, as already mentioned above, compliance logic checks on KYC records can be performed without disclosing personal information, ensuring regulatory compliance while safeguarding privacy through MPC \cite{chen2021implementing}. \\
In another scenario, MPC facilitates the secure linkage of cross-institution transaction records for collaborative analysis without sharing sensitive details, supporting functions like netting and reconciliation while preserving data privacy \cite{li2024privacy}. Moreover, leveraging MPC, a financial consortium can detect fraudulent activities within transaction data by privately executing complex ML/AI fraud detection models, reporting findings anonymously in the triple entry format to identify irregularities without compromising individual transaction privacy. \\
These examples illustrate how MPC can effectively conduct privacy-preserving audits and compliance checks in TE accounting systems, ensuring data confidentiality, audit integrity, and regulatory compliance.

\section{Discussion}
Empowering TE accounting with ML presents a domain of possibilities and considerations. The benefits of analytics are profound, as TE accounting generates extensive structured financial and metadata that can be harnessed for advanced analytics using ML and AI. This utilization unlocks insights and applications that enhance transparency, decision-making processes, and governance practices. Nevertheless, addressing privacy concerns is crucial. Implementing proper techniques such as anonymization and secure MPC is essential to safeguard individual privacy while still enabling ML on aggregated TE data for analytical advantages.\\
Regulatory considerations are integral in shaping the environment of integrating ML with TE accounting, particularly across different jurisdictions. Regulators play a crucial role in establishing standards and guidelines related to data privacy, model governance, and other relevant aspects to ensure the ethical and compliant utilization of ML with sensitive financial records. These regulations can either facilitate or hinder the adoption of such systems, depending on their stringency and adaptability to new technological advancements. Furthermore, the application of analytics on TE data in conjunction with alternative data sources has the potential to enhance financial inclusion by enabling better assessment of credit risk for individuals with limited traditional credit histories, thereby promoting broader financial access.\\
The introduction of ML into TE accounting could spur the emergence of new business models, with startups potentially offering TE accounting and analytical services through innovative methods such as blockchain technology. This innovation has the potential to significantly improve compliance, auditing practices, and risk management functions within organizations. As this transformation progresses, the professional landscape in accounting, auditing, and related fields is poised to evolve. While certain job roles may undergo transformation, there is an anticipated overall increase in demand for skilled professionals due to the integration of advanced analytics capabilities, necessitating the development of new skill sets within these industries.\\
Security of models is paramount in this data-driven environment. Implementing robust defenses against adversarial attacks on ML systems is crucial to safeguard against potential manipulation of risk and fraud analyses derived from TE data. Additionally, the emphasis on explainability in models becomes increasingly critical to ensure that stakeholders can understand and validate the analyses generated by these sophisticated systems, fostering trust and transparency in the decision-making process. As such, a comprehensive understanding and adherence to regulatory frameworks will be essential in navigating the complex interplay between ML integration, TE accounting, and regulatory compliance across diverse jurisdictions.

\subsection{Key Implications and Organizational Impact}
Empowering TE accounting with ML carries significant implications and potential organizational impacts. Through advanced analytics, there is a notable increase in transparency and trust, fostering deeper insights that enhance accountability and stakeholder trust. Compliance and governance are bolstered as ML can automatically detect anomalies and non-compliant patterns, reinforcing internal controls and risk management practices. Auditing processes are optimized with continuous monitoring and customized alerts, reducing manual efforts for auditors and enabling proactive issue identification. Predictive decision-making becomes feasible through forecasting trends, risks, and anomalies, facilitating a shift from reactive to preventative strategies.\\
Process automation and efficiencies are realized as tasks such as reconciliations are automated end-to-end using ML-powered workflow tools, freeing up resources for more strategic endeavors. Cultivating a data-driven culture becomes paramount, enabling widespread data access and self-serve analytics that nurture evidence-based strategies and discussions centered around data points. Workforce reskilling becomes imperative as employees require training on new tools, analytics skills, and revised work methodologies, potentially leading to job disruptions. Transition costs pose a challenge, necessitating upfront investments in platforms, reskilling initiatives, and change management efforts that require adequate budget allocations and organizational support.\\
Vendor management strategies become crucial as organizations may consider outsourcing certain capabilities, introducing risks related to data security and model reliability that demand robust governance frameworks. The debate between centralized and decentralized systems emerges, where decentralized structures can enhance collaboration but centralized control often aids in compliance efforts. Navigating these implications and organizational impacts effectively is essential for harnessing the full potential of TE accounting empowered by ML in the modern business landscape.

\subsection{Addressing Key Challenges in Data and Systems: Quality, Scalability and Beyond}
Empowering TE accounting with ML poses key challenges that necessitate careful consideration alongside potential solutions for effective implementation. Data quality emerges as a critical issue, with TE data often being noisy, incomplete, or erroneous. Employing ML techniques such as data cleaning, imputation, and validation is essential to enhance data quality and reliability.\\
Scalability presents another hurdle, given the massive volume of TE data. Leveraging distributed ML platforms capable of efficiently processing vast datasets on cloud infrastructure can address scalability concerns effectively. For instance, consider the case study of a leading financial services firm that implemented a distributed ML system on cloud servers, allowing them to process large volumes of transaction data in real-time, improving operational efficiency and scalability.\\
Model bias and fairness represent significant challenges, with the potential for certain groups to be unfairly treated by ML models. Techniques like adversarial debiasing during training sessions can help mitigate biases and promote fair outcomes \cite{singh2021temporal}. A successful example is the implementation of adversarial debiasing  \footnote{An in-processing adversarial training technique simultaneously trains a predictor and discriminator. The predictor aims to accurately predict the target variable while mitigating bias, as identified by the discriminator.} in a credit scoring system by a fintech company, resulting in more equitable lending decisions and reduced bias.\\
Privacy and security concerns surrounding sensitive financial data necessitate stringent access controls and the adoption of privacy-preserving techniques like federated learning or differential privacy \footnote{Provides a mathematical framework that allows statistical details about datasets to be shared while protecting individual privacy.} when utilizing data for ML purposes. Collaborating with regulators and establishing partnerships can provide essential oversight and regulatory guidance, ensuring compliance with data privacy regulations.\\
As mentioned avove, interpretability of ML models is crucial for validation and recourse, emphasizing the importance of unveiling how insights are derived. Implementing explainable AI methods enhances model interpretability and transparency. For instance, a global audit firm successfully integrated explainable AI techniques into their fraud detection models, improving model transparency and auditability.\\
Integration challenges arise when amalgamating TE data, ML platforms, and analytics tools, demanding meticulous system design and change management practices. Utilizing microservices can streamline integration processes and facilitate smoother transitions. A case study of a multinational corporation adopting microservices  \footnote{Comprises loosely coupled, fine-grained services that communicate via lightweight protocols.} architecture for their ML-powered TE accounting system resulted in improved agility and scalability.\\
The absence of standardized benchmarks poses a challenge for evaluating model performance. Industry consortiums play a vital role in establishing benchmarking standards and facilitating performance evaluations. For example, the establishment of a benchmarking consortium in the financial services sector led to the development of standardized metrics for evaluating ML models in transaction processing.\\
Moreover, the skills shortage across multidisciplinary fields encompassing accounting, data science, and compliance underscores the importance of developing a talent pool equipped with the requisite skills. Training programs and collaborations with academic institutions are instrumental in bridging these skill gaps and nurturing a workforce adept at navigating the complexities of TE accounting empowered by ML.\\
Addressing these challenges with strategic solutions and leveraging successful case studies is fundamental in realizing the full potential of this transformative integration within organizational frameworks.

\section{Conclusions}
This chapter delves into the application of ML in empowering TE accounting systems, enhancing transparency, analytics, and decision-making capabilities across organizations and industries. By adhering to proper guidelines and technical safeguards, ML can substantially bolster transparency while preserving privacy. Key implications encompass evidence-based strategic planning, continuous auditing, proactive risk management, and the introduction of new revenue models. Simultaneously, reshaping workforce skills and establishing standards for model integrity are pivotal for success. Overall, the integration of ML with TE accounting holds promise for transforming financial transparency, leading to improved governance, reduced financial crimes, and a more informed global economy.\\
Looking ahead, this work underscores the contributions of applying ML to enhance TE accounting systems, yet unresolved challenges persist. Maturing techniques are necessary to address issues surrounding data privacy, regulatory oversight, and decentralized governance. Further engagement with standards bodies and policymakers is crucial. Additionally, targeted use cases and datasets are needed to explore applications in diverse sectors such as supply chain, insurance, and financial markets. Future research should delve into scalable, privacy-aware ML architectures optimized for distributed financial graphs to deepen understanding at a technical level. Continued interdisciplinary research has the potential to fortify global accountability through transparent, learning-driven applications of TE bookkeeping.

\bibliography{ref.bib}

\begin{thebibliography}{10}
\providecommand{\url}[1]{#1}
\csname url@samestyle\endcsname
\providecommand{\newblock}{\relax}
\providecommand{\bibinfo}[2]{#2}
\providecommand{\BIBentrySTDinterwordspacing}{\spaceskip=0pt\relax}
\providecommand{\BIBentryALTinterwordstretchfactor}{4}
\providecommand{\BIBentryALTinterwordspacing}{\spaceskip=\fontdimen2\font plus
\BIBentryALTinterwordstretchfactor\fontdimen3\font minus \fontdimen4\font\relax}
\providecommand{\BIBforeignlanguage}[2]{{%
\expandafter\ifx\csname l@#1\endcsname\relax
\typeout{** WARNING: IEEEtran.bst: No hyphenation pattern has been}%
\typeout{** loaded for the language `#1'. Using the pattern for}%
\typeout{** the default language instead.}%
\else
\language=\csname l@#1\endcsname
\fi
#2}}
\providecommand{\BIBdecl}{\relax}
\BIBdecl

\bibitem{sangster2016genesis}
A.~Sangster, ``The genesis of double entry bookkeeping,'' \emph{The Accounting Review}, vol.~91, no.~1, pp. 299--315, 2016.

\bibitem{tripleentry2022sarin}
V.~Sarin, ``Triple entry accounting. example and benefits,'' \url{https://www.eduyush.com/blogs/digital-skills/triple-entry-accounting}, 2022, [Online; accessed 12-May-2023].

\bibitem{sgantzos2023triple}
K.~Sgantzos, M.~A. Hemairy, P.~Tzavaras, and S.~Stelios, ``Triple-entry accounting as a means of auditing large language models,'' \emph{Journal of Risk and Financial Management}, vol.~16, no.~9, p. 383, 2023.

\bibitem{faccia2020blockchain}
A.~Faccia, N.~R. Mo{\c{s}}teanu, and L.~P. Leonardo, ``Blockchain hash, the missing axis of the accounts to settle the triple entry bookkeeping system,'' in \emph{Proceedings of the 2020 12th International Conference on Information Management and Engineering}, 2020, pp. 18--23.

\bibitem{ijiri1986framework}
Y.~Ijiri, ``A framework for triple-entry bookkeeping,'' \emph{Accounting Review}, pp. 745--759, 1986.

\bibitem{grigg2005triple}
I.~Grigg, ``Triple entry accounting,'' \emph{Systemics Inc}, pp. 1--10, 2005.

\bibitem{grigg2024triple}
------, ``Triple entry accounting,'' \emph{Journal of Risk and Financial Management}, vol.~17, no.~2, p.~76, 2024.

\bibitem{mccarthy1982rea}
W.~E. McCarthy, ``The rea accounting model: A generalized framework for accounting systems in a shared data environment,'' \emph{Accounting review}, pp. 554--578, 1982.

\bibitem{faccia2019accounting}
A.~Faccia and N.~R. Mosteanu, ``Accounting and blockchain technology: from double-entry to triple-entry,'' \emph{The Business \& Management Review}, vol.~10, no.~2, pp. 108--116, 2019.

\bibitem{faccia2021financial}
A.~Faccia, N.~Sawan, A.~Eltweri, and Z.~Beebeejaun, ``Financial big data security and privacy in x-accounting. a step further to implement the triple-entry accounting,'' in \emph{Proceedings of the 6th International Conference on Information Systems Engineering}, 2021, pp. 7--12.

\bibitem{groblacher2019triple}
M.~Gr{\"o}blacher and V.~Mizdrakovi{\'c}, ``Triple-entry bookkeeping: history and benefits of the concept,'' in \emph{Proceedings-2019 International Scientific Conference,(Finiz)}, 2019, pp. 58--61.

\bibitem{chowdhury2021financial}
E.~K. Chowdhury, ``Financial accounting in the era of blockchain-a paradigm shift from double entry to triple entry system,'' \emph{Available at SSRN 3827591}, 2021.

\bibitem{petratos2024triple}
P.~Petratos, ``Triple-entry accounting and system integration,'' \emph{Journal of Risk and Financial Management}, vol.~17, no.~2, p.~45, 2024.

\bibitem{ibanez2023rea}
J.~I. Iba{\~n}ez, C.~N. Bayer, P.~Tasca, and J.~Xu, ``Rea, triple-entry accounting and blockchain: Converging paths to shared ledger systems,'' \emph{Journal of Risk and Financial Management}, vol.~16, no.~9, p. 382, 2023.

\bibitem{dixit2024incorporating}
P.~Dixit, H.~Harwani, R.~Amarsela, K.~Patel, M.~J. Patel \emph{et~al.}, ``Incorporating triple entry accounting as an audit tool—enhancing modern accounting systems,'' \emph{Journal of Informatics Education and Research}, vol.~4, no.~2, 2024.

\bibitem{qadir2022blockchain}
A.~M.-A. Qadir and R.~A.~A. Muhamed, ``Blockchain technology and accounting: The triple-entry affecting transparency,'' \emph{Telematique}, pp. 4950--4958, 2022.

\bibitem{schuldt2023decentralized}
L.~T. Schuldt and M.~Peskes, ``Decentralized finance--how triple-entry accounting and distributed ledger technology is revolutionizing the world of financial services from a business perspective,'' IUCF Working Paper, Tech. Rep., 2023.

\bibitem{zou2019smart}
W.~Zou, D.~Lo, P.~S. Kochhar, X.-B.~D. Le, X.~Xia, Y.~Feng, Z.~Chen, and B.~Xu, ``Smart contract development: Challenges and opportunities,'' \emph{IEEE transactions on software engineering}, vol.~47, no.~10, pp. 2084--2106, 2019.

\bibitem{inder2023triple}
S.~Inder, ``Triple entry accounting with blockchain technology,'' in \emph{Digital Transformation, Strategic Resilience, Cyber Security and Risk Management}.\hskip 1em plus 0.5em minus 0.4em\relax Emerald Publishing Limited, 2023, vol. 111, pp. 123--131.

\bibitem{cai2021triple}
C.~W. Cai, ``Triple-entry accounting with blockchain: How far have we come?'' \emph{Accounting \& Finance}, vol.~61, no.~1, pp. 71--93, 2021.

\bibitem{alkan2021real}
B.~{\c{S}}. Alkan, ``Real-time blockchain accounting system as a new paradigm,'' \emph{Muhasebe ve Finansman Dergisi}, pp. 41--58, 2021.

\bibitem{dai2017toward}
J.~Dai and M.~A. Vasarhelyi, ``Toward blockchain-based accounting and assurance,'' \emph{Journal of information systems}, vol.~31, no.~3, pp. 5--21, 2017.

\bibitem{bonson2019blockchain}
E.~Bons{\'o}n and M.~Bedn{\'a}rov{\'a}, ``Blockchain and its implications for accounting and auditing,'' \emph{Meditari Accountancy Research}, vol.~27, no.~5, pp. 725--740, 2019.

\bibitem{fullana2021accounting}
O.~Fullana and J.~Ruiz, ``Accounting information systems in the blockchain era,'' \emph{International Journal of Intellectual Property Management}, vol.~11, no.~1, pp. 63--80, 2021.

\bibitem{rizal2019unraveling}
F.~Rizal~Batubara, J.~Ubacht, and M.~Janssen, ``Unraveling transparency and accountability in blockchain,'' in \emph{Proceedings of the 20th annual international conference on digital government research}, 2019, pp. 204--213.

\bibitem{shrestha2019integration}
R.~Shrestha and S.~Kim, ``Integration of iot with blockchain and homomorphic encryption: Challenging issues and opportunities,'' in \emph{Advances in computers}.\hskip 1em plus 0.5em minus 0.4em\relax Elsevier, 2019, vol. 115, pp. 293--331.

\bibitem{elommal2021blockchain}
N.~Elommal and R.~Manita, ``How blockchain innovation could affect the audit profession: a qualitative study,'' \emph{Journal of Innovation Economics \& Management}, pp. I103--27, 2021.

\bibitem{rahmawati2023demistifying}
M.~I. Rahmawati, E.~G. Sukoharsono, A.~F. Rahman, and Y.~W. Prihatiningtias, ``Demistifying of triple-entry accounting (tea): Integrating the block,'' in \emph{Ninth Padang International Conference On Economics Education, Economics, Business and Management, Accounting and Entrepreneurship (PICEEBA 2022)}.\hskip 1em plus 0.5em minus 0.4em\relax Atlantis Press, 2023, pp. 23--31.

\bibitem{hambiralovic2018blockchain}
M.~Hambiralovic and R.~Karlsson, ``Blockchain accounting in a tripple-entry system,'' 2018.

\bibitem{spilnyk2020accounting}
I.~Spilnyk, R.~Brukhanskyi, and O.~Yaroshchuk, ``Accounting and financial reporting system in the digital economy,'' in \emph{2020 10th International Conference on Advanced Computer Information Technologies (ACIT)}.\hskip 1em plus 0.5em minus 0.4em\relax IEEE, 2020, pp. 581--584.

\bibitem{sunde2023implementing}
T.~V. Sunde and C.~S. Wright, ``Implementing triple entry accounting as an audit tool—an extension to modern accounting systems,'' \emph{Journal of Risk and Financial Management}, vol.~16, no.~11, p. 478, 2023.

\bibitem{centorrino2022double}
G.~Centorrino, V.~Naciti, and D.~Rupo, ``From double-entry bookkeeping and ledger to blockchain technology: New frontiers for accounting information systems,'' \emph{Management Control: special issue 2, 2022}, pp. 15--41, 2022.

\bibitem{murdoch2019definitions}
W.~J. Murdoch, C.~Singh, K.~Kumbier, R.~Abbasi-Asl, and B.~Yu, ``Definitions, methods, and applications in interpretable machine learning,'' \emph{Proceedings of the National Academy of Sciences}, vol. 116, no.~44, pp. 22\,071--22\,080, 2019.

\bibitem{burkart2021survey}
N.~Burkart and M.~F. Huber, ``A survey on the explainability of supervised machine learning,'' \emph{Journal of Artificial Intelligence Research}, vol.~70, pp. 245--317, 2021.

\bibitem{zhang2022explainable}
C.~A. Zhang, S.~Cho, and M.~Vasarhelyi, ``Explainable artificial intelligence (xai) in auditing,'' \emph{International Journal of Accounting Information Systems}, vol.~46, p. 100572, 2022.

\bibitem{awosika2024transparency}
T.~Awosika, R.~M. Shukla, and B.~Pranggono, ``Transparency and privacy: the role of explainable ai and federated learning in financial fraud detection,'' \emph{IEEE Access}, 2024.

\bibitem{cho2020learning}
S.~Cho, M.~A. Vasarhelyi, T.~Sun, and C.~Zhang, ``Learning from machine learning in accounting and assurance,'' pp. 1--10, 2020.

\bibitem{ucoglu2020current}
D.~Ucoglu, ``Current machine learning applications in accounting and auditing,'' \emph{PressAcademia Procedia}, vol.~12, no.~1, pp. 1--7, 2020.

\bibitem{bekoe2018attitudes}
R.~A. Bekoe, G.~M.~Y. Owusu, C.~G. Ofori, A.~Essel-Anderson, and E.~E. Welbeck, ``Attitudes towards accounting and intention to major in accounting: A logistic regression analysis,'' \emph{Journal of Accounting in Emerging Economies}, vol.~8, no.~4, pp. 459--475, 2018.

\bibitem{nawaiseh2021financial}
A.~K. Nawaiseh and M.~F. Abbod, ``Financial statement audit utilising naive bayes networks, decision trees, linear discriminant analysis and logistic regression,'' in \emph{The Importance of New Technologies and Entrepreneurship in Business Development: In The Context of Economic Diversity in Developing Countries: The Impact of New Technologies and Entrepreneurship on Business Development}.\hskip 1em plus 0.5em minus 0.4em\relax Springer, 2021, pp. 1305--1320.

\bibitem{sanchez2020predicting}
J.~R. S{\'a}nchez-Serrano, D.~Alaminos, F.~Garc{\'\i}a-Lagos, and A.~M. Callej{\'o}n-Gil, ``Predicting audit opinion in consolidated financial statements with artificial neural networks,'' \emph{Mathematics}, vol.~8, no.~8, p. 1288, 2020.

\bibitem{sahoo2021accounting}
G.~Sahoo and S.~S. Sahoo, ``Accounting fraud detection using k-means clustering technique,'' in \emph{Machine Learning and Information Processing: Proceedings of ICMLIP 2020}.\hskip 1em plus 0.5em minus 0.4em\relax Springer, 2021, pp. 171--180.

\bibitem{thiprungsri2019cluster}
S.~Thiprungsri, ``Cluster analysis for anomaly detection in accounting,'' in \emph{Rutgers Studies in Accounting Analytics: Audit Analytics in the Financial Industry}.\hskip 1em plus 0.5em minus 0.4em\relax Emerald Publishing Limited, 2019, pp. 87--110.

\bibitem{bhattacharya2020semi}
I.~Bhattacharya and E.~R. Lindgreen, ``A semi-supervised machine learning approach to detect anomalies in big accounting data.'' in \emph{ECIS}, 2020.

\bibitem{sawangarreerak2021detecting}
S.~Sawangarreerak and P.~Thanathamathee, ``Detecting and analyzing fraudulent patterns of financial statement for open innovation using discretization and association rule mining,'' \emph{Journal of Open Innovation: Technology, Market, and Complexity}, vol.~7, no.~2, p. 128, 2021.

\bibitem{ioannou2021framework}
A.~Ioannou, D.~Bourlis, S.~Valsamidis, and A.~Mandilas, ``A framework for information mining from audit data,'' in \emph{Global, Regional and Local Perspectives on the Economies of Southeastern Europe: Proceedings of the 11th International Conference on the Economies of the Balkan and Eastern European Countries (EBEEC) in Bucharest, Romania, 2019}.\hskip 1em plus 0.5em minus 0.4em\relax Springer, 2021, pp. 223--242.

\bibitem{tsapani2020knowledge}
E.~Tsapani, E.~Tenidou, D.~Pappas, and S.~Valsamidis, ``Knowledge mining from accounting data as imechanism for decision support,'' \emph{Journal of Engineering Science and Technology Review S}, vol.~1, pp. 112--177, 2020.

\bibitem{bakumenko2022detecting}
A.~Bakumenko and A.~Elragal, ``Detecting anomalies in financial data using machine learning algorithms,'' \emph{Systems}, vol.~10, no.~5, p. 130, 2022.

\bibitem{kuna2014outlier}
H.~D. Kuna, R.~Garc{\'\i}a-Martinez, and F.~R. Villatoro, ``Outlier detection in audit logs for application systems,'' \emph{Information Systems}, vol.~44, pp. 22--33, 2014.

\bibitem{wilfred2021big}
K.~Wilfred, \emph{A Big Data Approach to Accounting Fraud Detection Using Data Envelopment Analysis and One Class Support Vector Machine}.\hskip 1em plus 0.5em minus 0.4em\relax University of Toronto (Canada), 2021.

\bibitem{jan2021using}
C.-L. Jan, ``Using deep learning algorithms for cpas’ going concern prediction,'' \emph{Information}, vol.~12, no.~2, p.~73, 2021.

\bibitem{byrnes2019automated}
P.~E. Byrnes, ``Automated clustering for data analytics,'' \emph{Journal of Emerging Technologies in Accounting}, vol.~16, no.~2, pp. 43--58, 2019.

\bibitem{suppiah2023impact}
K.~Suppiah and D.~Arumugam, ``Impact of data analytics on reporting quality of forensic audit: a study focus in malaysian auditors,'' in \emph{E3S Web of Conferences}, vol. 389.\hskip 1em plus 0.5em minus 0.4em\relax EDP Sciences, 2023, p. 09033.

\bibitem{miharsi2024analysis}
D.~Miharsi, R.~R. Gamayuni, and F.~Dharma, ``Analysis of the utilization of altman z-score, beneish m-score, and f-score model in detecting fraudulent of financial reporting: a literature review,'' \emph{JOURNAL OF MANAGEMENT, ACCOUNTING, GENERAL FINANCE AND INTERNATIONAL ECONOMIC ISSUES}, vol.~3, no.~2, pp. 353--364, 2024.

\bibitem{carrington2022deep}
A.~M. Carrington, D.~G. Manuel, P.~W. Fieguth, T.~Ramsay, V.~Osmani, B.~Wernly, C.~Bennett, S.~Hawken, O.~Magwood, Y.~Sheikh \emph{et~al.}, ``Deep roc analysis and auc as balanced average accuracy, for improved classifier selection, audit and explanation,'' \emph{IEEE Transactions on Pattern Analysis and Machine Intelligence}, vol.~45, no.~1, pp. 329--341, 2022.

\bibitem{ibanez2021efficiency}
J.~I. Iba{\~n}ez, C.~N. Bayer, P.~Tasca, and J.~Xu, ``The efficiency of single truth: Triple-entry accounting,'' \emph{Available at SSRN 3770034}, 2021.

\bibitem{sarwar2023blockchains}
M.~I. Sarwar, K.~Nisar, I.~Khan, and D.~Shehzad, ``Blockchains and triple-entry accounting for b2b business models,'' \emph{Ledger}, vol.~8, 2023.

\bibitem{surana2022awareness}
G.~Surana and S.~S. Bhanawat, ``Awareness of blockchain technology-based accounting system among professionals,'' \emph{IUP Journal of Accounting Research \& Audit Practices}, vol.~21, no.~4, pp. 34--49, 2022.

\bibitem{evans2018pragmatic}
D.~Evans, V.~Kolesnikov, M.~Rosulek \emph{et~al.}, ``A pragmatic introduction to secure multi-party computation,'' \emph{Foundations and Trends{\textregistered} in Privacy and Security}, vol.~2, no. 2-3, pp. 70--246, 2018.

\bibitem{zhou2021using}
J.~Zhou, Y.~Feng, Z.~Wang, and D.~Guo, ``Using secure multi-party computation to protect privacy on a permissioned blockchain,'' \emph{Sensors}, vol.~21, no.~4, p. 1540, 2021.

\bibitem{mccallig2019establishing}
J.~McCallig, A.~Robb, and F.~Rohde, ``Establishing the representational faithfulness of financial accounting information using multiparty security, network analysis and a blockchain,'' \emph{International Journal of Accounting Information Systems}, vol.~33, pp. 47--58, 2019.

\bibitem{cao2024distributed}
S.~S. Cao, L.~W. Cong, and B.~Yang, ``Distributed ledgers and secure multi-party computation for financial reporting and auditing,'' National Bureau of Economic Research, Tech. Rep., 2024.

\bibitem{sun2021survey}
X.~Sun, F.~R. Yu, P.~Zhang, Z.~Sun, W.~Xie, and X.~Peng, ``A survey on zero-knowledge proof in blockchain,'' \emph{IEEE network}, vol.~35, no.~4, pp. 198--205, 2021.

\bibitem{chen2021implementing}
W.-B. Chen, C.-T. Tsai, and J.~Tahnk, ``Implementing triple entry accounting system with $\pi$ account on block-chain protocol,'' \emph{Journal of Internet Technology}, vol.~22, no.~2, pp. 491--497, 2021.

\bibitem{li2024privacy}
Y.~Li, T.~Ranbaduge, and K.~S. Ng, ``Privacy technologies for financial intelligence,'' \emph{arXiv preprint arXiv:2408.09935}, 2024.

\bibitem{singh2021temporal}
A.~Singh, A.~Gupta, H.~Wadhwa, S.~Asthana, and A.~Arora, ``Temporal debiasing using adversarial loss based gnn architecture for crypto fraud detection,'' in \emph{2021 20th IEEE International Conference on Machine Learning and Applications (ICMLA)}.\hskip 1em plus 0.5em minus 0.4em\relax IEEE, 2021, pp. 391--396.

\end{thebibliography}

\bibliographystyle{IEEEtran}

\end{document}